\documentstyle[epsf,12pt]{article}

\catcode`@=11 \@addtoreset{equation}{section} \catcode`@=12

\def\strutdepth{\dp\strutbox}
\def\marginarrow#1{\vtop to\strutdepth{
    \baselineskip\strutdepth\vss\llap{#1$\rightarrow$ }\null}}
\def\revision#1{\strut\vadjust{\kern-\strutdepth\marginarrow{#1}}}

\begin{document}
\begin{titlepage}
\begin{flushright}\vbox{\begin{tabular}{c}
           TIFR/TH/98-01\\
           January, 1998\\
           hep-ph/9801240\\
\end{tabular}}\end{flushright}
\begin{center}
   {\large \bf
      Quarkonium Polarization in Non-relativistic QCD\\
      and the Quark-Gluon Plasma}
\end{center}
\bigskip
\begin{center}
   {Sourendu Gupta\footnote{E-mail: sgupta@theory.tifr.res.in}\\
    Theory Group, Tata Institute of Fundamental Research,\\
    Homi Bhabha Road, Bombay 400005, India.}
\end{center}
\bigskip
\begin{abstract}
We examine angular distributions of leptons arising from the
decay of $J/\psi$ in inclusive hadroproduction. Taking into
account feed-down contributions from $\chi_{1,2}$, a flat
distribution emerges without fine tuning parameters. Changes
in the ratio of direct to total $J/\psi$ cross sections would
change this distribution significantly. Such angular distributions
are, therefore, confirmatory tests of the $J/\psi$ suppression
signal for the production of a quark-gluon plasma. Related
effects are predicted for the $\Upsilon$.
\end{abstract}
\end{titlepage}

\section{\label{intro}Introduction}

In the hadroproduction of $J/\psi$, the angular distribution of the
dileptons coming from the decay of the quarkonium is of the form
\begin{equation}
   {d\sigma\over d\cos\theta}\;\sim\;1+\alpha\cos^2\theta
\label{in.angle}\end{equation}
where $\theta$ is the angle (measured in the rest frame of the
quarkonium) between the direction of motion of the positively charged
lepton and the (lab-frame) direction of the quarkonium momentum.
As long as the initial particles are unpolarized, there can be no
difference in the weights for production of quarkonia with helicities
$m$ and $-m$. The angular distributions can then be summarized by an
``alignment parameter'' \cite{rose}---
\begin{equation}
   \zeta\;=\;{\sigma_L\over\sigma},\qquad{\rm and}\qquad
   \alpha\;=\;{1-3\zeta\over1+\zeta}.
\label{in.polm}\end{equation}
Here $\sigma_L$ is the cross section for production
of quarkonia with longitudinal polarization ($J_z=0$), and $\sigma$ is the
total cross section. For hadroproduction, $\alpha$ is close to zero for
all beam and target combinations investigated \cite{e672b,exps}, implying
that $\zeta$ is close to $1/3$.

The question of angular distributions and alignment has attracted a lot
of attention recently. In \cite{cho} it was pointed out that heavy-quark
spin symmetry has strong implications for $\zeta$. Non-relativistic QCD
(NRQCD) \cite{bbl}, which incorporates heavy-quark spin symmetry, is a
framework for understanding production rates of quarkonia. Polarization
in hadroproduction has been investigated in NRQCD in \cite{br1,br2,bchen}.
The dependence of $\alpha$ on kinematics is claimed to be a discriminant
between different models for hadroproduction of quarkonia, and hence a
crucial test of NRQCD \cite{br3}. Polarization of quarkonia produced in
the decay of $Z_0$ has also been investigated \cite{baek}, and may be
tested at LEP.

The phenomenological application of NRQCD to inclusive hadroproduction of
$J/\psi$ has been quite successful. Absolute cross sections are fully
compatible with the scaling laws of NRQCD, and the energy dependence of
total cross sections \cite{br1,short}, as well as longitudinal momentum
distributions \cite{long} are reproduced rather well. The ratio of $\chi_1$
and $\chi_2$ cross sections can also be understood by going to higher than
leading orders in the NRQCD expansion parameter, $v$ \cite{hiv}.
In view of this, the alignment of quarkonia produced in inclusive
hadron-hadron collisions is investigated. Although the computations are
performed to leading order in $v^2$ for each channel, this already
leads to an expansion to order $v^9$ in some cases.

Our results are summarized here. The observation of almost vanishing
$\alpha$ in inclusive $J/\psi$ production arises from the mixture of
direct $J/\psi$ and feed-down from $\chi$ states. We compute alignments
in NRQCD at sufficiently high order, thus supplementing the results of
\cite{br2,bchen} by formul\ae{} for $\zeta_\chi$. The value of $\alpha$
depends only on the ratios of certain non-perturbative matrix elements,
and is well constrained by the scaling laws of NRQCD. A flat angular
distribution emerges because of the feed-down from $\chi$ states.

This has implications for the $J/\psi$ suppression signal of the
quark-gluon plasma \cite{matsui}. A plasma melts not only the $J/\psi$,
the more massive quarkonium states are disrupted even more
easily--- indeed it has been speculated that some of them dissolve
even at temperatures lower than the QCD phase transition \cite{karsch}.
If this is so, then a suppression of the $J/\psi$ cross section due
to a plasma must be coupled with a large change in $\alpha$. This
provides a confirmatory test for the $J/\psi$ suppression signal of
the formation of a quark-gluon plasma in relativistic heavy-ion
collisions.

In section \ref{sc.nrqcd} the computation of the angular distribution in
NRQCD is outlined in brief. Details of the computation are given
in appendix \ref{sc.comp}, where a helicity technique used in conjunction
with previously known methods of computation is documented. In section
\ref{sc.res} the results for $\zeta$ are summarized. In the final section
\ref{sc.pheno} these results are applied to inclusive hadroproduction
and the signal of quark-gluon plasma formation.

\section{\label{sc.nrqcd} Non-relativistic QCD}

NRQCD is a low-energy effective theory for quarkonia \cite{bbl}. The
action is written in terms of all possible operators consistent with the
symmetries of QCD. All momenta in NRQCD are cut off by some scale $\Lambda$,
taken to be of the order of the heavy quark mass, $m$. The coupling associated
with each term is identified through a perturbative matching procedure. In
NRQCD the perturbative short-distance ($<1/\Lambda$) physics of the production
of a heavy quark pair ($\bar QQ$) is factored from the non-perturbative
long-distance physics of its hadronisation to a quarkonium. A proof of
factorisation has been given for production at large transverse momenta
\cite{bbl} and successful phenomenology has been done \cite{jpsi}. An
explicit test of factorisation for inclusive cross sections is provided by
a recent next-to-leading order, $\alpha_{\scriptscriptstyle S}^2$,
computation \cite{petr}.

The NRQCD factorisation formula for the inclusive production of heavy
quarkonium resonances $H$ with 4-momentum $P$ can be written as
\begin{equation}
   d\sigma\;=\;
      {\displaystyle{1\over\Phi}{d^3P\over(2\pi)^3 2E_{\scriptscriptstyle P}}}
     \sum_{ij} C_{ij}
          \left\langle{\cal K}_i\Pi(H){\cal K}^\dagger_j\right\rangle,
\label{nrqcd.nrqcd}\end{equation}
where $\Phi$ is a flux factor. The coefficient functions $C_{ij}$ are
computable in perturbative QCD and hence have an expansion in the strong
coupling $\alpha_{\scriptscriptstyle S}/\pi$ (evaluated at $\Lambda$).
Although each matrix element in the sum above is non-perturbative, it has
a fixed scaling dimension in the quark velocity $v$. Then the NRQCD cross
section is a double series in $\alpha_{\scriptscriptstyle S}/\pi$ and $v^2$.
For charmonium a numerical coincidence, $\alpha_{\scriptscriptstyle S}/\pi
<v^2$, makes it necessary to consider higher orders in $v^2$ before going
to higher orders in $\alpha_{\scriptscriptstyle S}$.

The fermion bilinear operators ${\cal K}_i$ are built out of heavy quark
fields sandwiching colour and spin matrices and the covariant derivative
${\bf D}$. The composite labels $i$ and $j$ include the colour index $\alpha$,
the spin quantum number $S$, the number of derivative operators $N$, the
orbital angular momentum $L$, the total angular momentum $J$, the helicity
$J_z$ and the parity. The hadron projection operator 
\begin{equation}
   \Pi(H)\;=\;{\sum_s} \left|H,s\rangle\langle H,s\right|,
\label{nrqcd.hproj}\end{equation}
(where $s$ denotes hadron states with energy less than the NRQCD cutoff),
is diagonal in appropriate bases \cite{cho,bbl,br2}. When the final state
helicities are summed it is diagonal in $J$, $J_z$ and parity. When the
hadron helicity is observed, heavy quark spin symmetry makes it diagonal
(up to corrections of order $v^2$) in the basis $\{S,L,J_z\}$.

The $J_z$-dependence of these matrix elements can be factored out using the
Wigner-Eckart theorem---
\begin{equation}\begin{array}{rl}
   &\langle{\cal K}_i\Pi(H){\cal K}^\dagger_i\rangle\;=\;
      {\displaystyle{1\over\hat J^2}}{\cal O}^H_\alpha({}^{2S+1}L_J^N),\\
   &\langle{\cal K}_i\Pi(H){\cal K}^\dagger_j\rangle+{\rm h.c.}\;=\;
      {\displaystyle{1\over\hat J\hat J'}}
         {\cal P}^H_\alpha({}^{2S+1}L_J^N,{}^{2S+1}L_{J'}^{N'}).
\end{array}\label{nrqcd.pred}\end{equation}
The factors of $\hat J=\sqrt{2J+1}$ come from Clebsch-Gordan coefficients
and are conventionally included in the coefficient function. NRQCD
supports a power counting rule for each matrix element---
\begin{equation}
   \langle{\cal K}_i\Pi(H){\cal K}^\dagger_j\rangle\;=\;R_H\Lambda^Dv^d,
\label{nrqcd.scale}\end{equation}
where $D$ is the mass dimension of the operator,
and the NRQCD order, $d$, is given by the rule
\begin{equation}
   d\;=\;3+N+N'+2(E_d+2M_d).
\label{nrqcd.rule}\end{equation}
$E_d$ and $M_d$ are the number of colour electric and magnetic transitions
required to connect the hadronic state to the state ${\cal K}_i|0\rangle$.
At tree level the dimensionless number $R_H$ in eq.\ (\ref{nrqcd.scale})
must depend only on the hadron $H$ under consideration. At higher orders
in $\alpha_{\scriptscriptstyle S}$, $R_H$ may be corrected by logarithms
of $\Lambda$.

\begin{table}[bh]\begin{center}\begin{tabular}{|c|c|c|c|}
\hline
Matrix Elements & Values & $R_H$ \\
\hline
${\cal O}^{J/\psi}_8\left({}^3S_1\right)$ &
                  $(1.12\pm0.14)\times10^{-2} {\rm\ GeV}^3$ & $0.22\pm0.03$ \\
${\cal O}^{J/\psi}_8\left({}^1S_0\right)+
            {7\over2m^2}{\cal O}^{J/\psi}_8\left({}^3P_0\right)$ &
                  $(3.90\pm1.14)\times10^{-2} {\rm\ GeV}^3$ & $0.17\pm0.05$\\
\hline
${\cal O}^{\psi'}_8\left({}^3S_1\right)$ &
                  $(0.46\pm0.08)\times10^{-2} {\rm\ GeV}^3$ & $0.09\pm0.02$\\
${\cal O}^{\psi'}_8\left({}^1S_0\right)+
            {7\over2m^2}{\cal O}^{\psi'}_8\left({}^3P_0\right)$ &
                  $(1.60\pm0.51)\times10^{-2} {\rm\ GeV}^3$ & $0.07\pm0.02$\\
\hline
\end{tabular}\end{center}
\caption[dummy]{The values of various non-perturbative matrix elements
  in the non-relativistic normalisation \cite{bkr}. These can be converted
  to the relativistic normalisation by multiplying by $4m$. The dimensionless
  number $R_H$ (eq.\ \ref{nrqcd.scale}) has been extracted using $\Lambda
  =m=1.5$ GeV and $v^2=0.3$.}
\label{tb.scale}\end{table}

In principle, we would like to do phenomenology with experimental
measurements of the matrix elements needed. However, the large number of
different matrix elements required for fixed target phenomena makes this
impossible at present. Consequently, we are forced to use the scaling laws
of eq.\ (\ref{nrqcd.scale}). Any number obtained in this way can at best
be indicative, and has to be justified by more detailed numerical work
when better data becomes available. The first step is, of course, to test
eq.\ (\ref{nrqcd.scale}). From the data summarized in Table \ref{tb.scale}
it seems that tree-level NRQCD scaling can be accepted as a working hypothesis.

The coefficient functions, $C_{ij}$, are computed using a non-relativistic
decomposition of heavy quark spinors and a Taylor expansion of the matrix
elements, $\cal M$, in the relative momentum, $q$, of the $\bar QQ$ pair
\cite{bchen}. Spherical tensor techniques are then used to recouple the
2-component spinor bilinear operators \cite{hiv}. The projection to specific
$J_z$ components can be performed at the matrix element level by appropriate
choice of gauge. The process of matching the perturbatively computed
$|{\cal M}|^2$ to the NRQCD formula of eq.\ (\ref{nrqcd.nrqcd}) is then simple.

\section{\label{sc.res}Quarkonium alignment}

We need to compute the alignment parameter only for $J/\psi$, $\chi_1$
and $\chi_2$. For completeness, we also list it for the yet unestablished
${}^1P_1$ state $h_c$. A simple extension of the results in \cite{bchen}
is sufficient to show that $\zeta^H_{\bar qq}=0$ for all quarkonia $H$ to
all orders in $v$. The total cross section  has been listed in \cite{hiv}
to order $\alpha_{\scriptscriptstyle S}^2v^9$. In this section only the
longitudinal parts of the $gg\to \bar QQ$ cross sections are listed. Some
details are given in Appendix \ref{sc.comp}.

\subsection{Direct $J/\psi$ alignment}
The direct $J/\psi$ subprocess longitudinal cross section is
\begin{equation}
   \hat\sigma^{J/\psi}_{gg}(\hat s) \;=\;
          \varphi \left[{5\over144}\theta^{J/\psi}_D(7)
                  +\left\{{5\over144}\theta^{J/\psi}_D(9)
                        +{1\over16}\theta^{J/\psi}_F(9)\right\}
		\right]
\label{res.jpsi}\end{equation}
where
\begin{equation}
\varphi = {\pi^3\alpha_s^2\over4m^2}\delta(\hat s-4 m^2),
\label{res.varphi}\end{equation}
and $\theta^{J/\psi}_a(n)$ denotes combinations of non-perturbative
matrix elements from the colour amplitude $a$ ($=S$, $D$ or $F$) at
order $v^n$. These can be written as
\begin{equation}\begin{array}{rl}
   \theta^{J/\psi}_D(7)\;=\;&
        {\displaystyle{1\over2m^2}}{\cal O}^{J/\psi}_8({}^1S_0^0)
       +{\displaystyle{3\over2m^4}}{\cal O}^{J/\psi}_8({}^3P_0^1)\\

   \theta^{J/\psi}_D(9)\;=\;&
       -{\displaystyle{23\over8\sqrt3m^4}}
                    {\cal P}^{J/\psi}_8({}^1S_0^0,{}^1S_0^2)
       -{\displaystyle{11\over8m^6}}\sqrt{\displaystyle{5\over3}}
                    {\cal P}^{J/\psi}_8({}^3P_0^1,{}^3P_0^3)
      \\&\qquad
          +{\displaystyle{2\over5m^6}}\sqrt{\displaystyle{2\over3}}
                    {\cal P}^{J/\psi}_8({}^3P_0^1,{}^3P_2^3)\\

   \theta^{J/\psi}_F(9)\;=\;&
        {\displaystyle{1\over2m^6}}{\cal O}^{J/\psi}_8({}^3P^2_1)\\
\end{array}\label{res.jpsime}\end{equation}

The $\theta^{J/\psi}_D(7)$ term agrees with that given in \cite{bchen}.
The rest of the terms did not appear there because the Taylor expansion
was truncated at lower order. Terms involving ${}^1P_1$ operators, given
in \cite{bchen}, contribute at order $v^{11}$ and hence are not included in
our results.

The leading, order $v^7$, terms have been considered before \cite{br1,br2}.
Under the assumption of tree-level NRQCD scaling this gives
\begin{equation}
    \zeta^{J/\psi}_{gg} \;=\; {1\over6} + {\cal O}(v^2).
\label{res.jpsieta}\end{equation}
Order $v^2$ corrections come from two sources--- the terms in the production
cross section given above, as well as from corrections to the NRQCD action
through terms that violate heavy-quark spin symmetry.

\subsection{$\chi_1$ alignment}
The longitudinal cross section for $\chi_1$ is
\begin{equation}
   \hat\sigma^{\chi_1}_{gg}(\hat s) \;=\;
		\varphi \left[{1\over54}
        \theta^{\chi_1}_S(9) +{5\over144} \theta^{\chi_1}_D(9) 
        +{1\over16} \theta^{\chi_1}_F(9)
		     \right]
\label{res.chi1}\end{equation}
with the combinations
\begin{equation}\begin{array}{rl}
   \theta^{\chi_1}_S(9)\;=\;&
        {\displaystyle{3\over2m^4}}{\cal O}^{\chi_1}_1({}^3P_0^1)\\

   \theta^{\chi_1}_D(9)\;=\;&
        {\displaystyle{1\over2m^2}}{\cal O}^{\chi_1}_8({}^1S_0^0)
       +{\displaystyle{3\over2m^4}}{\cal O}^{\chi_1}_8({}^3P_0^1)\\

   \theta^{\chi_1}_F(9)\;=\;&
        {\displaystyle{1\over6m^4}}{\cal O}^{\chi_1}_8({}^1P^1_1)
       +{\displaystyle{1\over18m^6}}{\cal O}^{\chi_1}_8({}^3S^2_1)
       +{\displaystyle{5\over18m^6}}{\cal O}^{\chi_1}_8({}^3D^2_1).
\end{array}\label{res.chi1me}\end{equation}
Although $\chi_1$ production begins at order $v^9$, the large number of
terms involved makes its cross section comparable to that of $\chi_{0,2}$.
Alignment of $\chi_1$ has not been considered before in the literature.
Under the assumption of tree-level NRQCD scaling we have
\begin{equation}
    \zeta^{\chi_1}_{gg} \;=\; {55\over239} + {\cal O}(v^2).
\label{res.chi1eta}\end{equation}

\subsection{$\chi_2$ alignment}

The leading term in the total cross section for $\chi_2$ production
is given by a $J_z=\pm2$ operator scaling as $v^5$. The most significant
$J_z=0$ operator scales as $v^7$---
\begin{equation}
   \hat\sigma^{\chi_2}_{gg}(\hat s) \;=\;\varphi
       \displaystyle{1\over90}\theta^{\chi_2}_S(7),
      \qquad{\rm and}\qquad
   \theta^{\chi_2}_S(7)\;=\;
        {\displaystyle{5\over2\sqrt3m^4}}
                {\cal P}^{\chi_2}_1({}^3P_0^1,{}^3P_0^3)
\label{res.chi2me}\end{equation}
Since the total cross section starts at order $v^5$, the alignment is
\begin{equation}
    \zeta^{\chi_1}_{gg} \;=\; 0 + {\cal O}(v^2).
\label{res.chi2eta}\end{equation}

\subsection{$h_c$ alignment}
The production cross section for the ${}^1P_1$ charmonium state is---
\begin{equation}
   \hat\sigma^{h_c}_{gg}(\hat s) \;=\; \varphi{\displaystyle{5\over144}}
       \biggl[\theta^{h_c}_D(5)+\theta^{h_c}_D(7)\biggr],
\label{res.hc}\end{equation}
where the combinations of non-perturbative matrix elements can be
written as
\begin{equation}
   \theta^{h_c}_D(5)\;=\;
        {\displaystyle{1\over2m^2}}{\cal O}^{h_c}_8({}^1S_0^0)
      \qquad{\rm and}\qquad
   \theta^{h_c}_D(7)\;=\;
       -{\displaystyle{23\over8\sqrt3m^4}}
                {\cal P}^{h_c}_8({}^1S_0^0,{}^1S_0^2).
\label{res.hcme}\end{equation}
In this case,
\begin{equation}
   \zeta^{h_c}_{gg}={1\over3} + {\cal O}(v^2),
\label{res.hceta}\end{equation}
implying that angular distributions are trivial.

\section{\label{sc.pheno}Phenomenology}

In computing the effective alignment of $J/\psi$, the feed-down from
radiative decays $\chi_J\to\gamma J/\psi$ must be taken into account.
The total spin-projected cross section for $J/\psi$ can be written as
\begin{equation}
   \sigma^{tot}_{J/\psi}(J_z)\;=\;\sigma_{J/\psi}(J_z)
      +\sum_J B_J \sum_M p^J_{J_z,M}\sigma_{\chi_J}(M),
\label{pheno.eff}\end{equation}
where $B_J$ denotes the branching ratio for the decay of
$\chi_J$ to $J/\psi$, and $B_Jp^J_{J_z,M}$ is the
branching fraction for a $\chi_J$ with spin projection $M$ to give a
$J/\psi$ with spin projection $J_z$. We use the notation
\begin{equation}
   f_J\;=\;{B_J\sigma_{\chi_J}
          \over\sigma_{J/\psi}+B_1\sigma_{\chi_1}+B_2\sigma_{\chi_2}},
\label{pheno.note}\end{equation}
to write the effective alignment parameter as
\begin{equation}
   \zeta\;=\;\zeta_{J/\psi}(1-f_1-f_2)
       +f_1 \biggl[p^1_{00}\zeta_{\chi_1}+P^1_{01}(1-\zeta_{\chi_1})\biggr]
       +f_2 P^2_{02}.
\label{pheno.form}\end{equation}
In writing the equation above, the fact that $\chi_2$ is produced only with
$J_z=\pm2$ at leading order in $v$ is taken into account, and the notation
$P^J_{0,J}=p^J_{0,J}+p^J_{0,-J}$ is introduced. Observe
that the first term on the right has the form of a dilution factor over
$\zeta_{J/\psi}$, driving the effective $\zeta$ away from the required value
of $1/3$. This has to be compensated by effects of the other two terms.

The cross section for the production of any charmonium
state is obtained by convoluting the parton level cross sections with
parton luminosity factors---
\begin{equation}
   \sigma_H(J_z;\sqrt S)\;=\;
     {\cal L}_{\bar qq}(\sqrt S)\hat\sigma^H_{\bar qq}(J_z)
       +{\cal L}_{gg}(\sqrt S)\hat\sigma^H_{gg}(J_z),
\label{pheno.conv}\end{equation}
where $\sqrt S$ is the centre of mass energy at which the cross section
is measured. Recalling that the alignment from the $\bar qq\to\bar QQ$
channels is identically zero, it is clear that
\begin{equation}
   \zeta_H\;=\;\zeta^H_{gg}\left[1+
        {{\cal L}_{\bar qq}\over{\cal L}_{gg}}
        {\hat\sigma^H_{\bar qq}\over\hat\sigma^H_{gg}}\right]^{-1}.
\label{pheno.true}\end{equation}
The second factor is a dilution. The ratio of parton level cross sections
on the right is determined by the NRQCD scaling laws. As a result, this
dilution factor is completely determined once the parton densities are
specified. In the rest of this paper the GRV LO parton densities are used
for both proton and pion \cite{grv}.

\begin{figure}
\epsfbox{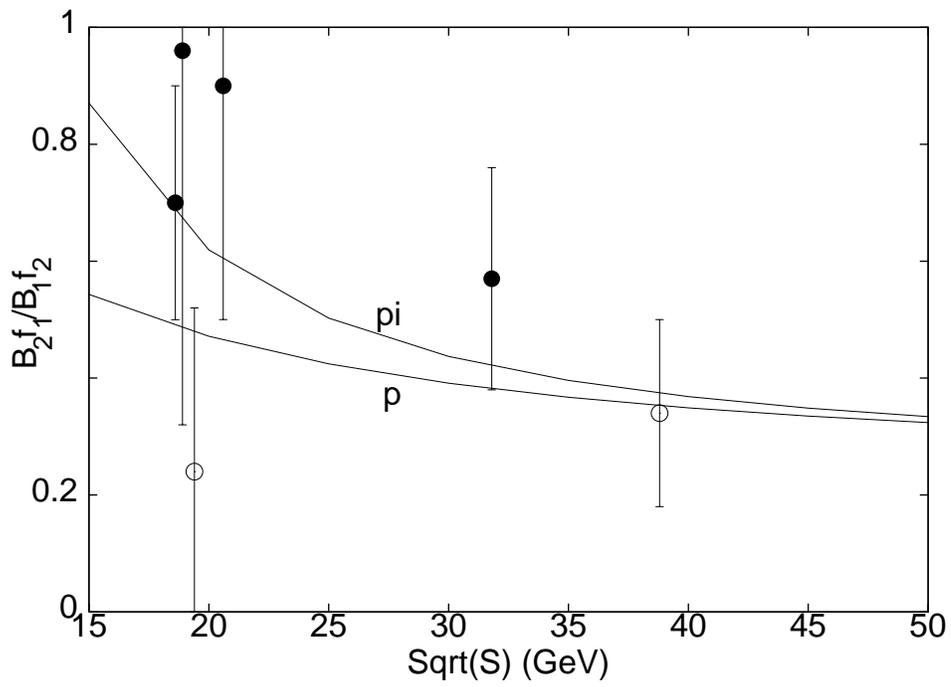}
\caption[dummy]{Measurements of $B_2f_1/B_1f_2$ (collected in \cite{br3})
   in $\pi$p (filled circles) and pp collisions (open circles) are compared
   with the predictions of NRQCD.}
\label{fg.f1f2}\end{figure}

The quantities $f_1$ and $f_2$ have been investigated in several
experiments. Pion beams have been used at many energies, and
there is weak evidence for a non-trivial $\sqrt S$ dependence.
The number of measurements with proton beams is more limited. Some of the
older experiments, at low $\sqrt S$, \cite{old} had reported $f_1/f_2$
somewhat smaller than those observed with pion beams. A new measurement
at higher $\sqrt S$ \cite{e771} seems to give a value consistent with the
pion data at a similar energy. The data has been collected in \cite{br3}.
Using the cross sections for $\chi$ production computed to order
$\alpha_{\scriptscriptstyle S}v^9$ \cite{hiv}, and tree level NRQCD scaling
(eq.\ \ref{nrqcd.scale}) with $R_{\chi_1}=R_{\chi_2}$, reasonable agreement
with data is obtained\footnote{For some colour singlet matrix elements, $R_H$
is related to the radial wavefunction of the hadron $H$. Although this
intuitive picture is violated in NRQCD, in particular by logarithms of
$\Lambda$, it provides the motivation for the choice $R_{\chi_1}=R_{\chi_2}$.
It would be preferable to use data, when it becomes available, to fix these
two parameters.}. As shown in Figure \ref{fg.f1f2}, NRQCD scaling implies
that the $f_1/f_2$ ratio should show significant beam dependence at small
$\sqrt S$ but should rapidly converge to a smaller common value with
increasing $\sqrt S$. This is due to the relatively large influence of
${\cal L}_{\bar qq}$ at small $\sqrt S$, specially for $\pi$-A collisions.
Since the main uncertainty is due to $\alpha_{\scriptscriptstyle S}$
corrections to eq.\ (\ref{nrqcd.scale}), we estimate a theoretical
uncertainty of about 25\% for predictions of $f_1/f_2$.

The next step is to constrain the three parameters $p^1_{00}$, $P^1_{01}$
and $P^2_{02}$. In the decay $\chi_J\to\gamma J/\psi$, the photon
energy is of the order of $0.5$ GeV, and much less than the NRQCD cutoff.
As a result, the decay matrix elements, and hence $p^J_{0,M}$, cannot be
computed in NRQCD through the usual factorisation approach. One way
out is to obtain them by direct measurement. Each of the LEP experiments
should have a few hundred identified $\chi$'s in the hadronic decay mode
of the $Z_0$. The parameters $p^J_{0,M}$ can be estimated from angular
correlations between the decay products in the chain
$\chi\to\gamma J/\psi\to\gamma e^+e^-$.

An alternative technique has been used in the literature \cite{tang}.
Assuming that the decay is dominated by an electric dipole transition,
using heavy-quark spin symmetry, and integrating over the direction of
the photon momentum, the parameters can be shown to be just squares of
certain Clebsch-Gordan coefficients---
\begin{equation}
   p^J_{J_z,M}\;=\;\left|C(1J1;M-J_z,M)\right|^2.
\label{pheno.tang}\end{equation}
Contributions of higher multipoles are expected to be subdominant, since
they go as powers of $E_\gamma/M_{J/\psi}$ ($E_\gamma$ is the energy of
the decay photon, and $M_{J/\psi}$ is the mass of the $J/\psi$). With
this model, $p^1_{00}=0$, $P^1_{01}=1$, $P^2_{02}=0$.

\begin{figure}
\epsfbox{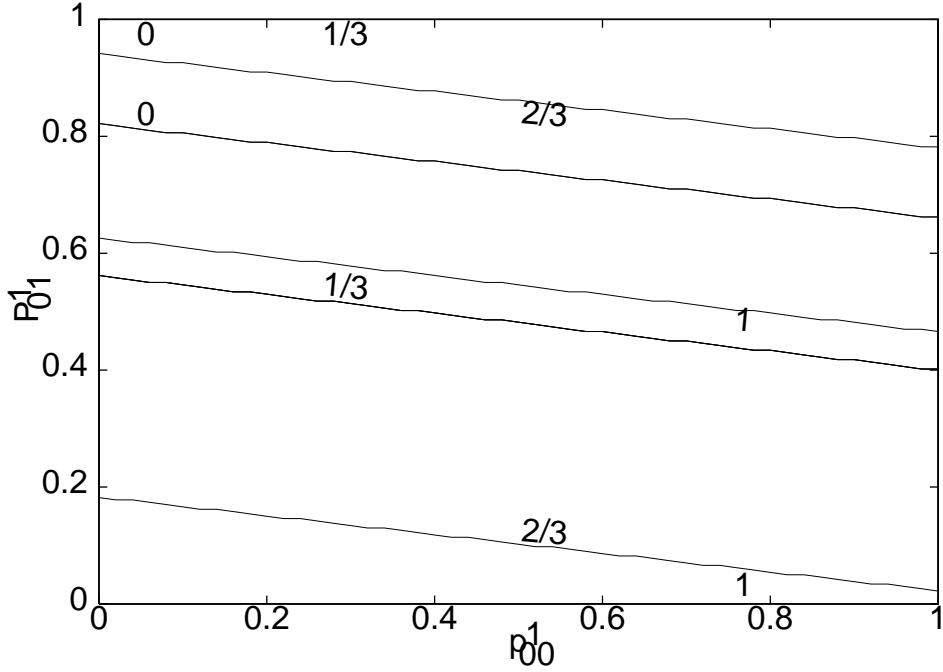}
\caption[dummy]{Contours of allowed values of $p^1_{00}$ and $P^1_{01}$. The
   upper and lower lines of the contour are indicated by marking them with
   the relevant value of $P^2_{02}$.}
\label{fg.excl}\end{figure}

In \cite{e672b} it is reported that in $\pi$-Be collisions at $\sqrt S=
31.8$ GeV,
\begin{equation}
   \alpha\;=\;-0.01\pm0.12,
\label{pheno.alpha}\end{equation}
implying $\zeta=0.34\pm0.05$. The same experimental collaboration also
reports \cite{e672a}
\begin{equation}
   f_1+f_2=0.443\pm0.041\pm0.035
     \quad{\rm and}\quad
   {B_2 f_1\over B_1 f_2}\;=\;0.57\pm0.18\pm0.06.
\label{pheno.fpi}\end{equation}
Extracting the values of $f_1$ and $f_2$ from these measurements, we
find that eq.\ (\ref{pheno.form}) and the assumption of dipole dominated
decay give
\begin{equation}
   \zeta=0.30\pm0.04\qquad{\rm and}\qquad\alpha=0.08\pm0.08.
\label{pheno.effalpha}\end{equation}
The errors in the prediction are obtained solely by propagation from
errors in $f_1$ and $f_2$ (statistical and systematic errors are added
in quadrature). We have assumed that the errors in $f_1$ and $f_2$ are
uncorrelated. The theoretical uncertainty from neglecting order $v^2$
corrections to $\zeta_{gg}$ and order $\alpha_{\scriptscriptstyle S}$
corrections to tree level NRQCD scaling is not displayed in
eq.\ (\ref{pheno.effalpha}).

We can also investigate the allowed departure from the dipole model by
taking $f_1$, $f_2$ and $\alpha$ from the above experiment, and finding
the contours of allowed $p^1_{00}$, $P^1_{01}$ and $P^2_{02}$ using
eq.\ (\ref{pheno.form}). The 1-$\sigma$ allowed contours are shown in
Figure \ref{fg.excl}. It is clear that the dipole model, as well as
large deviations from it, are allowed.

Similar results follow if we combine measurements of $\alpha$ and $f_{1,2}$
from different experiments with the same beam and roughly similar energy.
We can also make these estimates by combining the NRQCD scaling predictions
for $f_1/f_2$ with experimental measurements of $f_1+f_2$. The same features
emerge--- the dipole dominance model of the radiative decay of $\chi$ used
in conjuction with eq.\ (\ref{pheno.form}) works well, and departures from
it are allowed. In other words, strong tuning of parameters is not required
to reproduce the angular distribution of leptons from the decay of
hadroproduced $J/\psi$.

An application to $\bar bb$ systems may also be considered. In this case
it is estimated that $v^2\approx0.1$. As a result, hadroproduction of
$\Upsilon$ should have roughly equal contributions from the order $v^5$
production of $\chi_2$ (with $B_2=0.2$) and the order $v^7$ direct
production of $\Upsilon$. The other $\chi_b$ states should be much less
important--- the $\chi_0$ contribution being suppressed by the smaller
branching ratio, $B_0<0.06$, and the $\chi_1$ by the relative order $v^4$
in the matrix elements. Assuming therefore, $f_2^b=0.5$ and $f_1^b=0$,
eq.\ (\ref{pheno.form}) predicts $\alpha_\Upsilon\approx0.25$--0.70. The
larger value is preferred if $\chi_b$ decays occur through electric
dipole transitions.

\begin{figure}
\epsfbox{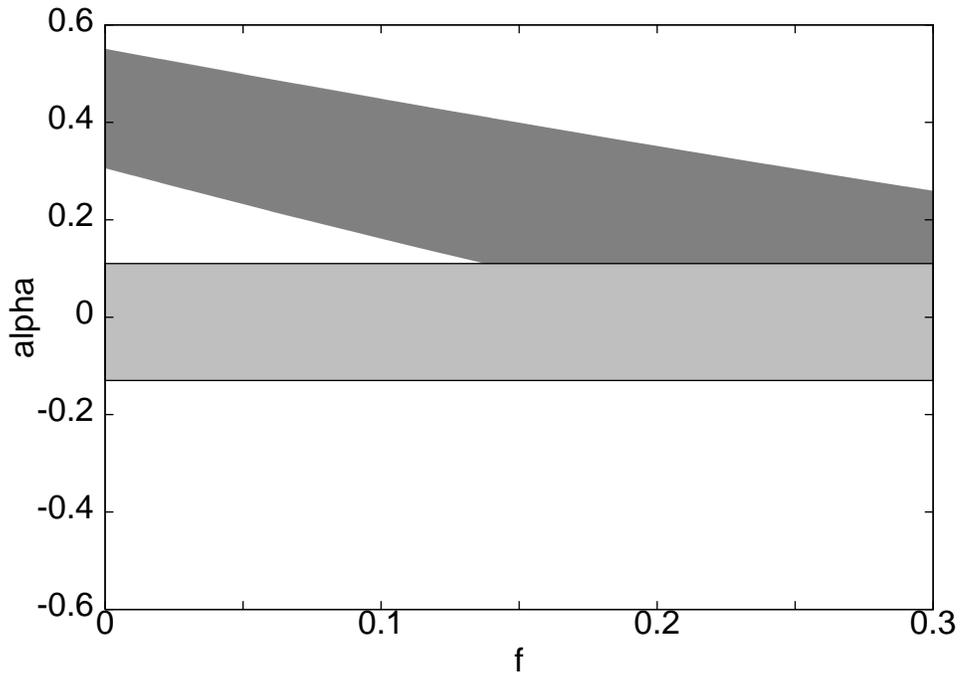}
\caption[dummy]{The parameter $\alpha$ as a function of the fraction of
   $J/\psi$ produced by decay of $\chi_{1,2}$. The dark band is the
   prediction for $f=f_1=f_2$. The lighter horizontal band is the
   1-$\sigma$ range for $\alpha$ seen in $T=0$ experiments.}
\label{fg.plasma}\end{figure}

Consider now nucleus-nucleus collisions where the central rapidity
region reaches a temperature sufficient to suppress $\chi$ but not
$J/\psi$ \cite{karsch}, {\sl i.\ e.\/}, $f_1(T)\approx f_2(T)\approx0$.
The total $J/\psi$ cross section should decrease by about 50\%.
Simultaneously the angular distribution should change, with a new
$\zeta=3/7$. As shown in Figure \ref{fg.plasma}, the change should
be easily visible.

Thermal effects on $\Upsilon$ have not been considered earlier, since
this resonance is not expected to dissolve at the temperatures which
the LHC may reach \cite{karsch}. However, the $\chi_b$ states are
expected to disappear at such temperatures \cite{karsch}. Our earlier
analysis would then lead us to believe that some reduction of the
$\Upsilon$ cross section may occur, and the angular distribution may
change to give $\alpha_\Upsilon=3/7$. Since the successive disappearance
of different quarkonium states can be used for thermometry of the plasma,
we believe it is important to look for this effect.

We present a list of experiments which might be used to constrain and test
the applicability of NRQCD to inclusive quarkonium hadroproduction.
\begin{itemize}
\item
   More refined measurements of the cross sections for $\chi_1$ and
   $\chi_2$ production with proton and pion beams, at $\sqrt S\approx
   40$ GeV, should test the scaling of the matrix elements and fix the
   ratio $R_{J/\psi}/R_{\chi}$. High-statistics experiments at $\sqrt
   S<25$ GeV designed to see the difference in $f_1/f_2$ for proton and
   pion would be welcome.
\item
   Measurement of angular correlations in the decay $\chi\to\gamma e^+e^-$
   would test the dipole dominance model of this decay, and hence the
   NRQCD explanation of the angular distribution. Hadroproduction
   experiments are not needed for this. In fact $e^+e^-$ machines, such
   as the LEP, would be preferable.
\item
   Present day errors on $\alpha$ are large, and need to be reduced
   significantly. Simultaneous measurements of $f_1$, $f_2$ and
   $\alpha$ by one experiment can test NRQCD scaling rather accurately.
\item
   The $\bar bb$ system remains almost completely unexplored. Due to the
   smallness of $v^2$ in this system, one would expect the phenomenology to
   be quite different from the $\bar cc$ system, and scaling arguments to
   work better. In view of the importance of $\Upsilon$ and $\chi_b$ for
   tests of NRQCD scaling, and possible applications to the thermometry
   of a quark-gluon plasma, a vigorous experimental effort in this field
   would pay large dividends.
\end{itemize}

The results can be summarized as follows--- an NRQCD based argument
yields $\alpha_{J/\psi}\approx0$, provided the scaling formula of
eq.\ (\ref{nrqcd.scale}) is used. The effective angular distribution
(eq.\ \ref{pheno.form}) needs three parameters related to angular
correlations in the decays $\chi_J\to\gamma J/\psi$. They need not
be fine tuned to reproduce the observed angular distributions,
although the assumption that the decays proceed through an electric
dipole transition is supported. A crude estimate for the $\bar bb$
system gives $\alpha_\Upsilon\approx0.25$--0.70. These findings may
be applied to the $J/\psi$ suppression signal for the formation of a
quark-gluon plasma--- a suppression should be coupled with a large
change in $\alpha$, since the feed-down processes disappear rapidly
with increasing temperature. Suppression of $\chi_b$ should similarly
influence the angular distribution of $\Upsilon$.

\newpage\appendix
\section{\label{sc.comp} Computing the coefficient functions}

We take the momenta of the initial particles, $p_1$ and $p_2$, and the
momentum of the pair, $P=p_1+p_2$ ($P^2=M^2$), to lie along the $z$-direction.
We choose $p_1^z>0$ and take this to be the axis of quantization of angular
momenta. The 4-momenta of $Q$ and $\bar Q$ ($p$ and $\bar p$
respectively, $p^2=\bar p^2=m^2$) are written as
\begin{equation}
   p\;=\;{1\over2}P+L_j q^j\qquad{\rm and}\qquad
   \bar p\;=\;{1\over2}P-L_j q^j.
\label{comp.momdef}\end{equation}
The space-like vector $q$ is defined in the rest frame of the pair,
and $L^\mu_j$ boosts it to any frame. For $gg$ initial states we choose
polarization vectors
\begin{equation}
   \epsilon^1(\lambda)\;=\;{1\over\sqrt2}(0,-\lambda,-i,0)
     \qquad{\rm and}\qquad
   \epsilon^2(\lambda)\;=\;{1\over\sqrt2}(0,\lambda,-i,0).
\label{comp.gauge}\end{equation}
Since the polarization vectors are orthogonal to both the initial
momenta, this corresponds to a choice of planar gauge.

Euclidean 3-tensors are expressed as spherical tensors. As an example,
a 3-vector $a_i$ is written as a spherical tensor of rank 1, with
components
\begin{equation}
   a_{\pm1}\;=\;\mp{1\over\sqrt2}\left(a_x\pm ia_y\right),\quad
   a_0\;=\;a_z.
\label{comp.spten}\end{equation}
In this representation
\begin{equation}
   \epsilon^1(\lambda)\cdot L_m\;=\;-\delta_{m,\lambda},
\label{comp.example}\end{equation}
where $m$ is a helicity index. An useful identity is
\begin{equation}
   a_j b_j\;=\; a_0 b_0 - (a_{+1}b_{-1} + a_{-1}b_{+1})
                                        \;=\;-\sqrt3[a,b]^0_0,
\label{comp.stidf}\end{equation}
We have introduced the notation $[a,b]^J_M$ to denote two vectors
$a$ and $b$ coupled to total rank $J$ and helicity $M$. The coefficient
of the terms $[a,b]^0_0$ can be obtained from the appropriate Clebsch-Gordan
coefficients.

Since we work to lowest order in the QCD coupling, the perturbative
projector is
\begin{equation}
   \Pi(\bar QQ)\;=\;|\bar QQ\rangle\langle\bar QQ|.
\label{comp.qqbproj}\end{equation}
Its normalisation is completely fixed by the relativistic normalisation
of states
\begin{equation}
   \langle Q(p,\xi)\bar Q(\bar p,\eta)|Q(p',\xi')\bar Q(\bar p',\eta')\rangle
     \;=\;4 E_p E_{\bar p} (2\pi)^6\delta^3(p-p')\delta^3(\bar p-\bar p'),
\label{comp.norm}\end{equation}
where the spinor normalisations are $\xi^\dagger\xi=\eta^\dagger\eta=1$.
Expanding $E_p=E_{\bar p}=\sqrt{m^2+q^2}$ in $q^2$ allows us to write any
spinor bilinear in terms of transition operators built out of heavy quark
fields.

The matrix element for the subprocess $\bar qq\to\bar QQ$ has been treated
to leading order in $v$ in \cite{bchen}. The result, $\zeta_{\bar qq}=0$
for all quarkonium states, can be trivially generalized to all orders in $v$.
The squared matrix element can be written in the form
\begin{equation}
   |{\cal M}|^2 \;=\; \left[k_1\cdot L_i k_2\cdot L_j
      + k_2\cdot L_i k_1\cdot L_j - k_1\cdot k_2 L_i\cdot L_j\right]
       W^{ij},
\label{comp.qqbar}\end{equation}
where $W^{ij}$ represents the square of a heavy-quark spinor bilinear. In
the kinematics appropriate for this problem, the prefactor to $W^{ij}$
becomes $(1-\delta_{i3}\delta_{j3})$. Transforming to spherical tensor
components, this can be seen to give a vanishing longitudinal cross section.

The computations for the $gg$ process can easily become tedious. We introduce
here a helicity technique based on a decomposition of the matrix element into
gauge invariant amplitudes---
\begin{equation}
   {\cal M}\;=\;{1\over6}g^2\delta_{ab}S
               +{1\over2}g^2d_{abc}D^c
               +{i\over2}g^2f_{abc}F^c.
\label{comp.ampl}\end{equation}
These amplitudes have been enumerated earlier \cite{hiv}.

The invariant amplitudes are most compactly written in terms of two
quantities $\cal A$ and $\cal S$. When the initial gluon helicities
are $\lambda$ and $\lambda'$, with the gauge choice given earlier,
we get
\begin{equation}
  {\cal A}\;=\;{1\over2}i\lambda\delta_{\lambda,\lambda'}
     \quad{\rm and}\quad
   {\cal S}_{ij}a^ib^j\;=\;
         -\sqrt3\left[a,b\right]^0_0\delta_{\lambda,\lambda'}
         +\sqrt2\left[a,b\right]^2_{2\lambda}\delta_{\lambda,-\lambda'}
\label{comp.not4}\end{equation}
In order to identify all terms to order $v^9$ we need the
colour amplitude $S$ to order $q^5$---
\begin{equation}\begin{array}{rl}
   S\;&=\;-\left({\displaystyle8im\over\displaystyle M}\right){\cal A}
                   \,(\xi^\dagger\eta)
       + {\displaystyle 4\over\displaystyle M} {\cal S}_{jm}
                   \,(q^m\xi^\dagger\sigma^j\eta)
       - \left({\displaystyle32im\over\displaystyle M^3}\right){\cal A}
                   \hat z_m\hat z_n\,(q^mq^n\xi^\dagger\eta)
     \\ & \qquad\quad
       + {\displaystyle16\over\displaystyle M^3}
           \left[{\cal S}_{jm}\hat z_n\hat z_p
                     -{\displaystyle M\over\displaystyle M+2m}\delta_{jm}
                {\cal S}_{np}\right]\,(q^mq^nq^p\xi^\dagger\sigma^j\eta)
     \\ & \qquad\qquad
       - \left({\displaystyle128im\over\displaystyle M^5}\right){\cal A}
                   \hat z_m\hat z_n\hat z_p\hat z_r
             \,(q^mq^nq^pq^r\xi^\dagger\eta)
     \\ & \quad\quad
       + {\displaystyle64\over\displaystyle M^5}
           \left[{\cal S}_{jm}\hat z_n\hat z_p
                     -{\displaystyle M\over\displaystyle M+2m}\delta_{jm}
                {\cal S}_{np}\right]\hat z_r\hat z_s
                     \,(q^mq^nq^pq^rq^s\xi^\dagger\sigma^j\eta).
\end{array}\label{comp.ampls}\end{equation}
The amplitude $D$ differs only through having colour octet matrix elements
in place of the colour singlet ones shown above. For the colour amplitude
$F$ we need the expansion
\begin{equation}
   F^c\;=\;-\left({\displaystyle16im\over\displaystyle M^2}\right){\cal A}
                   \hat z_m\,(q^m\xi^\dagger T^c\eta)
       + {\displaystyle8\over\displaystyle M^2}{\cal S}_{jm}\hat z_n
                   \,(q^mq^n\xi^\dagger\sigma^jT^c\eta).
\label{comp.amplf}\end{equation}
In all three colour amplitudes, the terms in ${\cal A}$ are spin singlet and
those in ${\cal S}$ are spin triplet.

The structure is very simple. Conservation of angular momentum shows that
the $J_z=0$ terms are obtained when $\lambda=\lambda'$. With eq.\ 
(\ref{comp.not4}) recoupling of all spherical tensors into terms of
well-defined $L$, $S$ and $J$ can now be performed at the amplitude level.
This also simplifies the computations presented in \cite{hiv}. The full
procedure, from this recoupling to the computation of the coefficient
functions can be reduced to a Mathematica program.


\newpage
\begin{thebibliography}{99}
\bibitem{rose}
   M.\ E.\ Rose, {\sl ``Elementary Theory of Angular Momentum''\/},
    John Wiley and Sons, New York, 1957.
\bibitem{e672b}
   A.\ Gribushin {\sl et al.\/}, {\sl Phys.\ Rev.\/}, D 53 (1996) 4723.
\bibitem{exps}
   J.\ G.\ Heinrich {\sl et al.\/}, {\sl Phys.\ Rev.\/} D 44 (1991) 1909;\\
   C.\ Akerlof {\sl et al.\/}, {\sl Phys.\ Rev.\/} D 48 (1993) 5064;\\
   T.\ Alexopoulos {\sl et al.\/}, {\sl Phys.\ Rev.\/} D 55 (1997) 3927.
\bibitem{cho}
   P.\ Cho and M.\ Wise, {\sl Phys.\ Lett.\/}, B 346 (1995) 129.
\bibitem{bbl}
   W.E.\ Caswell and G.P.\ Lepage, {\sl Phys.\ Lett.\/}, B 167 (1986) 437;\\
   G.\ T.\ Bodwin, E.\ Braaten and G.\ P.\ Lepage,
   {\sl Phys.\ Rev.\/}, D 51 (1995) 1125;
   [Erratum {\sl ibid.\/}, D 55 (1997) 5853].
\bibitem{br1}
   M.\ Beneke and I.\ Rothstein, {\sl Phys.\ Lett.\/}, B 372 (1996) 137;
   [Erratum: {\sl ibid.\/} B 389 (1996) 789].
\bibitem{br2}
   M.\ Beneke and I.\ Rothstein, {\sl Phys.\ Rev.\/}, D 54 (1996) 2005;
   [Erratum: {\sl ibid.\/} D 54 (1996) 7082].
\bibitem{bchen}
   E.\ Braaten and Y.\ Chen, {\sl Phys.\ Rev.\/}, D 54 (1996) 3216.
\bibitem{br3}
   M.\ Beneke, preprint hep-ph/9703429, to appear in the Proceedings
     of the XXIVth SLAC Summer Institute on Particle
     Physics, August 1996.
\bibitem{baek}
   S.\ Baek, P.\ Ko, J.\ Lee and H.\ S.\ Song, {\sl Phys.\ Rev.\/},
     D 55 (1997) 6839.
\bibitem{short}
   S.\ Gupta and K.\ Sridhar, {\sl Phys.\ Rev.\/}, D 54 (1996) 5545.
\bibitem{long}
   S.\ Gupta and K.\ Sridhar, {\sl Phys.\ Rev.\/}, D 55 (1997) 2650.
\bibitem{hiv}
   S.\ Gupta and P.\ Mathews, {\sl Phys.\ Rev.\/}, D 56 (1997) 3019;\\
   S.\ Gupta and P.\ Mathews, {\sl Phys.\ Rev.\/}, D 56 (1997) 7341.
\bibitem{matsui}
   T.\ Matsui and H.\ Satz, {\sl Phys.\ Lett.\/}, B 178 (1986) 416.
\bibitem{karsch}
   F.\ Karsch and H.\ Satz, {\sl Z.\ Phys.\/}, C 51 (1991) 209.
\bibitem{jpsi}
   E.\ Braaten, M.\ A.\ Doncheski, S.\ Fleming and M.\ Mangano,
     {\sl Phys.\ Lett.\/}, B 333 (1994) 548;\\
   D.\ P.\ Roy and K.\ Sridhar, {\sl Phys.\ Lett.\/}, B 339 (1994) 141;\\
   M.\ Cacciari and M.\ Greco, {\sl Phys.\ Rev.\ Lett.\/}, 73 (1994) 1586. 
\bibitem{petr}
   A.\ Petrelli, M.\ Cacciari, M.\ Greco, F.\ Maltoni and M.\ L.\ Mangano,
      preprint hep-ph/9707223.
\bibitem{bkr}
   M.\ Beneke and M.\ Kramer, {\sl Phys.\ Rev.\/}, D 55 (1997) 5269.
\bibitem{grv}
   M. Gluck, E.\ Reya and A.\ Vogt, {\sl Z.\ Phys.\/}, C 67 (1995) 433;\\
   M. Gluck, E.\ Reya and A.\ Vogt, {\sl Z.\ Phys.\/}, C 53 (1992) 651.
\bibitem{old}
   D.\ A.\ Bauer, {\sl et al.\/}, {\sl Phys.\ Rev.\ Lett.\/}, 54 (1985) 753;\\
   L.\ Antoniazzi {\sl et al.\/}, {\sl Phys.\ Rev.\/}, D 49 (1994) 543.
\bibitem{e771}
   K.\ Hagan {\sl et al.\/} (E 771), to appear in the Proceedings of the
     Quarkonium Physics Workshop, University of Illinois, Chicago, June
     1996; measurements quoted in \cite{br3}.
\bibitem{tang}
   M.\ V\"anttinen, P.\ Hoyer, S.\ J.\ Brodsky and W.-K.\ Tang,
      {\sl Phys.\ Rev.\/}, D 51 (1995) 3332;\\
   W.-K.\ Tang and M.\ V\"anttinen, {\sl Phys.\ Rev.\/}, D 54 (1996) 4349.
\bibitem{e672a}
   V.\ Koreshev {\sl et al.\/}, {\sl Phys.\ Rev.\ Lett.\/}, 77 (1996) 4294.
\end{thebibliography}
\end{document}